\documentclass[hyper,letterpaper,12pt,oneside]{JHEP3}
\usepackage{amsmath,amssymb}
\usepackage{bm}
\usepackage{cite}

\newcommand{\gsim}{\mathrel{\lower2.5pt\vbox{\lineskip=0pt\baselineskip=0pt
            \hbox{$>$}\hbox{$\sim$}}}}
\newcommand{\lsim}{\mathrel{\lower2.5pt\vbox{\lineskip=0pt\baselineskip=0pt
            \hbox{$<$}\hbox{$\sim$}}}}

\newcommand{\vev}[1]{\left\langle#1\right\rangle}

\newcommand{\order}[1]{{\cal O}\left( #1 \right)}
\newcommand{\e}{\text{e}}
\newcommand{\ev}{\text{eV}}
\newcommand{\mev}{\text{MeV}}
\newcommand{\gev}{\text{GeV}}

\newcommand{\kev}{\text{keV}}
\newcommand{\beq}{\begin{eqnarray}}
\newcommand{\eeq}{\end{eqnarray}}
\newcommand{\nn}{\nonumber}
\newcommand{\ie}{{\it i.e.}}

\newcommand{\hc}{\text{h.c.}}
\newcommand{\Mpl}{M_{\rm P\ell}}

\preprint{BUHEP-04-05\\ \hepph{0405083}} 

\title{Searching for Composite Neutrinos in the Cosmic Microwave Background}

\author{Takemichi Okui\\
Physics Department, Boston University,\\
Boston, MA 02215, USA}

\abstract{We analyze signals in the Cosmic Microwave Background (CMB) in 
theories where the small Dirac neutrino masses arise as a consequence of the 
compositeness of right-handed neutrinos.  In such theories, the right-handed 
neutrinos are massless ``baryons'' of a new strong gauge interaction.  We find 
that the results crucially depend on whether or not the new strong sector 
undergoes chiral symmetry breaking.  In the case with chiral symmetry breaking, 
we find that there are indeed signals in the CMB, but none of them is a direct 
consequence of neutrino compositeness.  In contrast, if the underlying theory 
does not undergo chiral symmetry breaking, the large scattering cross-section 
among the composites gives rise to a sizable CMB signal over a wide region of 
the parameter space, and it can potentially probe whether the neutrino mass 
spectrum is hierarchical, inverse-hierarchical, or degenerate.  We also 
discuss collider constraints on the compositeness in the context of the CMB 
signals.}


\begin{document}

\section{Introduction}
\label{sec:intro}
While a considerable amount of data has accumulated over the past several 
years, neutrinos still remain to be one of the most mysterious parts of the 
standard model of particle physics.  Is the neutrino mass spectrum 
hierarchical, inverse hierarchical, or degenerate?  Are neutrino masses 
Dirac or Majorana?  What is the underlying mechanism that generates these 
incredibly tiny masses---at least six orders of magnitude smaller than the 
electron mass---and large mixing angles? 

Recently, signals in the Cosmic Microwave Background (CMB) that can probe 
these questions are proposed by Z.~Chacko, L.~Hall, T.O. and S.~Oliver 
\cite{Chacko:2003dt}.  They investigate signals in theories where small 
neutrino masses arise from flavor symmetry spontaneously broken at a 
very low energy scale.  The CMB signals arise because such a theory necessarily 
introduces new particles interacting with neutrinos at sufficiently low 
energies to generate small neutrino masses.  Consequently, properties of 
the neutrino fluid during the acoustic oscillation at $T\approx\ev$ can be 
significantly modified, which in turn leaves signals in the CMB.  For example, 
an increase in the energy density of the neutrino fluid will appear in CMB 
observation as a deviation of the effective number of neutrino species from 
the standard value, namely, three.  A change in the scattering property of the 
neutrino fluid can also affect the CMB; while in the standard cosmology 
neutrinos are assumed to be 
free-streaming during the generation of the CMB at $T\approx\ev$, scattering 
of neutrinos with the new particles can prevent neutrinos from free-streaming.  
This effect will appear in the CMB in a very dramatic way---all the acoustic 
peaks of the CMB spectrum will shift {\it uniformly} by the amount 
$\Delta \ell \approx 8$ for each {\it non}-free-streaming neutrino species. 

It should be clear, however, that the existence of these signals is 
qualitatively quite generic to {\it any} theory where small neutrino masses 
are generated at sufficiently low energies, because whatever sector that 
generates the neutrino masses must involve new particles and new interactions 
with neutrinos, and if the associated energy scale is sufficiently low, it can 
affect the CMB spectrum. 

In fact, for Dirac neutrinos, another mechanism for naturally generating 
small masses is known; neutrino masses are tiny because the right-handed 
neutrinos are composite \cite{Arkani-Hamed:1998pf}.  One theoretical aspect 
where this mechanism is clearly more attractive than generating small neutrino 
masses from low-energy flavor symmetry breaking is that there is no issue of 
stabilizing a very low VEV required for obtaining small neutrino masses. 
In a flavor symmetry breaking scenario we need to add some extra ingredient, 
such as supersymmetry, into the theory to ensure the stability, but as far as 
the CMB signals are concerned, those extra complications are completely 
irrelevant.  In sharp contrast, the stability of the scale in a composite 
neutrino scenario is automatically guaranteed thanks to dimensional 
transmutation.  While experimental bounds on the compositeness of the charged 
fermions have been pushed higher and higher in energy, it is quite interesting 
that there is large unexplored room for the compositeness of neutrinos, which 
can be exploited to explain mysteries like small neutrino masses, and which 
can be tested by looking at the sky.  

In this paper we explore questions like: how large are the CMB signals for 
the composite neutrinos?   Can we tell from the CMB signals whether or not  
the strong sector that produces the composite right-handed neutrinos breaks 
chiral symmetry dynamically?  Can we distinguish a composite-neutrino scenario 
from a flavor-symmetry-breaking scenario?  In section \ref{sec:fatnu}, we will 
review the idea of composite neutrinos introduced in 
Ref.~\cite{Arkani-Hamed:1998pf}.  In section \ref{sec:cmbsignals}, we will 
review the CMB signals proposed in Ref.~\cite{Chacko:2003dt}, emphasizing the 
model-independent features relevant for us.  Section \ref{sec:region} is the 
main part of the paper in which we will analyze the CMB signals for two 
qualitatively different scenarios, one with chiral symmetry breaking and the 
other without.  We will show that a very low compositeness scale is necessary 
for the CMB signals to be generated.  In section \ref{sec:collider} we will 
discuss constraints from terrestrial experiments on such a low compositeness 
scale.  We will conclude in section \ref{sec:conc}.  In the appendix, we 
present two {\it concrete} models that produce exactly three massless fermionic 
composites with and without chiral symmetry breaking.  Although our discussions
of the CMB signals are completely model-independent, it is quite satisfactory 
to have actual, concrete models that do what we want.

\section{Neutrinos Are Light Because They Are Fat!}
\label{sec:fatnu}
In this section we will briefly review how compositeness can naturally explain 
tiny neutrino masses \cite{Arkani-Hamed:1998pf}.  We exclusively deal with the 
Dirac type neutrinos.%
\footnote{One way to ensure the Dirac nature of neutrinos is to impose 
lepton-number (or $B-L$) conservation.  Another way is to break lepton-number 
(or $B-L$) at an extremely high scale; for example, a Planck-suppressed 
operator $\ell\ell hh/ \Mpl$, gives an utterly negligible contribution to 
neutrino masses, so the Dirac-type contributions from the compositeness as 
described in this section will completely dominate.
}
First, note that there is absolutely no difficulty in {\it describing} a small 
Dirac neutrino mass; we can just write down a Yukawa coupling, $\ell \nu_R h$, 
with an appropriately small coefficient.  (Here $\ell$ and $h$ are the 
standard-model lepton and Higgs doublets, respectively, while $\nu_R$ is the 
right-handed neutrino.)  However, since neutrino oscillation data 
\cite{Fukuda:1998mi, Ahmad:2002jz, Eguchi:2002dm, Smy:2003jf} suggest that the 
heaviest neutrino mass is about $0.05~\ev$ in the case of a hierarchical or 
inverse-hierarchical mass spectrum, we find that even the largest coefficient 
of this Yukawa coupling is about $3\times 10^{-13}$.  Although there are no 
theoretical (nor experimental) problems with this small coupling, especially 
given the fact that the electron Yukawa coupling is $\order{10^{-6}}$, it is 
small enough that it is worth attempting to explain the smallness.  The idea of 
{\it composite} (or {\it fat}) right-handed neutrinos explains quite 
elegantly why Dirac neutrinos can be so light.  

Suppose that there is a new sector with a new asymptotically-free gauge 
interaction which confines at a scale $\Lambda$, under which all the 
standard-model fields are neutral.  We call this new color force 
{\it nu-color}.  Also suppose that the nu-color sector only communicates with 
the neutrino sector of the standard model, via some irrelevant operators.  (No 
relevant or marginal operators are capable of connecting the two sectors.)  
More explicitly, suppose that, at some energy scale $M\gg\Lambda$, the 
following operator is generated:
\beq
   {\cal L} = \cdots + \frac{1}{M^{3(n-1)/2}} \ell h 
                       \!\!\underbrace{qq \cdots q}_{n~\text{``quarks''}}  
\label{eq:op-at-M} ~,
\eeq
where a $q$ schematically refers to a ``quark'' charged under the nu-color 
gauge group.  (In general there are different kinds of $q$'s, namely, not 
necessarily all the ``quarks'' are in the same representation of the nu-color 
group.  But since this sort of details are unimportant here, we just 
collectively call them $q$.)  Since the $q$'s are fermions, $n$ must be an odd 
integer. 

Now, at the scale $\Lambda$, nu-color dynamics confines and produces a bunch 
of ``baryons'', ``mesons'', ``glueballs'', etc.  Suppose that the confinement 
produces three massless ``baryons'' of the form corresponding to the 
combination $qq \cdots q$ in (\ref{eq:op-at-M}), whose masslessness is ensured 
by some unbroken chiral symmetry of the nu-color dynamics.%
\footnote{For explicit examples of such gauge theories, see Appendix.  The 
examples in the appendix are all 4D non-supersymmetric asymptotically-free 
gauge theories.  A different, interesting framework to realize composite 
neutrinos is to use the AdS/CFT duality, where the right-handed neutrinos are 
composites of the CFT \cite{Gherghetta:2003he}.
}
We identify these three massless ``baryons'' as the three right-handed 
neutrinos.  Then, beneath $\Lambda$, in terms of the canonically-normalized 
right-handed neutrino fields, $\nu_R \approx qq \cdots q / \Lambda^{3(n-1)/2}$, the operator (\ref{eq:op-at-M}) becomes
\beq
   {\cal L} = \cdots + \left( \frac{\Lambda}{M} \right)^{3(n-1)/2}  
                         \ell \nu_R h
\label{eq:op-at-Lambda} ~.
\eeq
If we assume for simplicity that the original operator (\ref{eq:op-at-M}) has 
an $\order{1}$ coefficient for the heaviest neutrino in the case of a 
hierarchical or inverse-hierarchical spectrum, then we obtain the following 
relation:
\beq
   \left( \frac{\Lambda}{M} \right)^{3(n-1)/2} \approx \frac{m_{\nu_3}}{v}  ~,
\label{eq:M-Lambda}
\eeq    
where $m_{\nu_3} \approx 0.05 \> \ev$, and $v$ is the electroweak VEV, 
$174\>\gev$.  Because of this power-law relation, a small hierarchy in 
$\Lambda/M$ can reproduce the large hierarchy in $m_{\nu_3}/v$.  

Let us study some numerics about $M$ and $\Lambda$.  For example, in the 
``minimal'' case of $n=3$, we have
\beq
   \Lambda \approx 0.7 \times 10^{-4} M 
                 ~~~\text{if $\nu_R \propto qqq$} ~~(n=3)~,
\label{eq:qqq}
\eeq
while for $n=5$ we have
\beq
   \Lambda \approx 0.8 \times 10^{-2} M 
                 ~~~\text{if $\nu_R \propto qqqqq$} ~~(n=5)~.
\label{eq:qqqqq}
\eeq
Even though $10^{-4}$ may still be regarded as a small number, it is an 
``acceptable'' small number because the electron Yukawa coupling, for example, 
is $\order{10^{-6}}$.  This is a significant improvement compared to the 
original hierarchy, $10^{-13}$.

Note that for the purpose of getting small neutrino masses, only the ratio of 
$\Lambda$ to $M$ matters.  However, in this paper we are interested in lowering
the entire scale down so that there are some effects on the CMB.  Then, what 
is the lower bound on $M$?  Clearly, we are safe if $M \gsim v$.  This 
actually turns out to be the case, as we will see in section 
\ref{subsec:chiral}, {\it if} the nu-color dynamics undergoes chiral symmetry 
breaking.  On the other hand, if $M$ is lower than $v$, it is not immediately 
clear whether this is excluded or not, because the answer depends on how the 
operator (\ref{eq:op-at-M}) is generated.  We will come back to this point 
later in detail after we analyze the case with{\it out} chiral symmetry 
breaking, where we will see that $\Lambda$ must be indeed very low to generate 
the CMB signals.

\section{The CMB Signals}
\label{sec:cmbsignals}
As discussed in \cite{Chacko:2003dt}, there are two qualitatively different 
signals that can probe how neutrino masses are generated at low energy.  
Here we will briefly review these signals, emphasizing model-independent, 
essential features of the signals. 

The most important event that must happen before anything else is 
thermalization among left-handed neutrinos, right-handed neutrinos and 
other particles from the nu-color sector.  Obviously, to generate any signal 
in the CMB this thermalization must occur before the temperature of the 
universe becomes $\order{\ev}$, since otherwise the evolution of the universe
would be completely standard all the way down to the era of the acoustic 
oscillation.  However, the thermalization should not take place too early 
either, because we know that Big-Bang Nucleosynthesis (BBN) works quite well 
in the standard cosmology, which in particular constrains the number of 
relativistic degrees of freedom (converted into the effective number of 
neutrino species) {\it during the BBN era} to be $2.4 \pm 0.8$ at $2\sigma$ 
level \cite{Barger:2003zg}.%
\footnote{Note that this is a constraint from the BBN data alone 
(\ie~no CMB or large scale structure data are used), which is appropriate 
for our purpose because we are modifying cosmology in the later era.}
This bound arises because the D and $^4$He 
abundances are very sensitive to the expansion rate of the universe which is 
proportional to the square-root of the number of relativistic degrees of 
freedom.  Therefore, we have room only for the standard three left-handed 
neutrinos, and it is only {\it after} BBN can we bring the particles from the 
nu-color sector into thermal equilibrium with left-handed neutrinos.  
Following \cite{Chacko:2003dt}, we call this thermalization event 
{\it neutrino-recoupling}, meaning that the left-handed neutrinos experience 
thermalization once again after they decoupled from the photon-baryon-electron
plasma.

\subsection{The $\Delta N_\nu$ Signal}
\label{subsec:deltaNnu}
This signal utilize the fact that the CMB spectrum is sensitive to the 
energy density of relativistic degrees
of freedom when the temperature of the universe is $\order{\ev}$.  If the 
energy density of relativistic degrees of freedom is increased, the equality 
of matter and radiation occurs later, so the time window between the equality 
and the photon-decoupling decreases.  Consequently, the sound-wave of the 
photon-baryon plasma has less time to travel, so the size of the sound-horizon
at the time of the photon-decoupling shrinks.  Since the location of the first
CMB peak is the ratio of the present horizon to the sound-horizon then (modulo
effects of curvature), we see that increasing the radiation energy density 
moves the first peak to the right (\ie~higher multipole $\ell$).  Furthermore,
since the gravitational potential due to dark matter has less time to grow 
before the photon-decoupling, the heights of the potential hills that photons 
must climb up become lower, so the CMB photons red-shifts less. Hence, the CMB
spectrum becomes more blue, \ie, the first CMB peak not only moves to the 
right but also becomes higher.  Therefore, the CMB spectrum is very sensitive 
to the radiation energy density at $T\approx\ev$ \cite{Hu:1995fq}.  It is 
customary in cosmology to express the extra (\ie~other than photons) 
relativistic degrees of freedom in terms of the effective number of neutrino 
species.  The current bounds from WMAP alone is $0.9\lsim N_\nu \lsim 8.3$ at 
the $2\sigma$ level \cite{Barger:2003zg}, while adding late time data such as 
large scale structure tightens the bound slightly to 
$1.9 \lsim N_\nu \lsim 7.0$ \cite{Hannestad:2003xv}, which still allows 
room for nonstandard $N_\nu$.     

Now we are ready to discuss how $N_\nu$ can be different from 3 in our 
scenario.  To illustrate the idea, imagine that in addition to our three 
composite $\nu_R$'s, the nu-color sector contains another light particle, 
$\phi$, which couples to neutrinos schematically as $g \,\nu_L \nu_R \phi$.
(In the flavor symmetry breaking scenarios discussed in 
Ref.~\cite{Chacko:2003dt}, $\phi$ is a pseudo-Goldstone boson from the flavor 
symmetry breaking.)  Let us begin the story from the moment right after BBN.  
Because of the BBN constraint on $N_\nu$ we discussed earlier, we are 
initially allowed to have 
only $\nu_L$'s in the system.  Now, the interaction $g \,\nu_L \nu_R \phi$ 
will bring $\nu_L$, $\nu_R$ and $\phi$ into equilibrium before $T$ becomes 
$\order{\ev}$ if $g$ is sufficiently large and the rate of the process 
$\nu_L\bar{\nu}_L \to \phi\phi^*$ decreases less rapidly than the expansion 
rate does.  Note that neutrino-recoupling requires $\phi$ to be relativistic, 
since a scattering process involving non-relativistic $\phi$ in the final 
state would have a highly suppressed phase space.  So after 
neutrino-recoupling we are left with a gas of relativistic $\nu_L$, $\nu_R$ 
and $\phi$.  

Now, if $m_\phi \lsim \ev$, we will just observe $N_\nu = 3$ in the CMB, 
because in this case the gas will stay relativistic all the way the down to 
the matter-radiation equality, and by energy conservation the total energy 
density of the $\nu_L$-$\nu_R$-$\phi$ gas will be the same as that of the 
pure $\nu_L$ gas in the standard cosmology as if nothing had occurred.  Since 
the CMB is sensitive only to the total energy density and not to the 
composition of the gas, there will be no observable effect in this case. 

On the other hand, if $m_\phi \gsim \ev$, the $\phi$ particles go 
non-relativistic before the matter-radiation equality.  Recall that a similar 
situation occurs to electrons when the temperature drops below the
electron mass.  In that case, electron-positron pairs annihilate into photons,
heating up the photon temperature by the famous $(11/4)^\frac13$ factor.  
Exactly the same thing happens in our case as well; pairs of $\phi$'s 
annihilate into neutrinos, raising the neutrino temperature.  This will 
appear in the CMB as $N_\nu >3$.  Clearly, the strength of this signal, or 
the amount of deviation of $N_\nu$ from 3, depends on how many neutrinos had 
recoupled to $\phi$ in the first place.  For example, in the scenarios 
discussed in Ref.~\cite{Chacko:2003dt}, the coupling $g$ is proportional to 
$m_\nu$, so a heavier neutrino is more likely to recouple than a lighter one, 
and this is why measuring $N_\nu$ from the CMB can tell us how many neutrino 
species are heavy, by which we can distinguish the three different possible 
mass spectra of neutrinos.  As we will see, an analogous 
situation occurs in our scenario as well.

\subsection{The $\Delta \ell$ Signal}
\label{subsec:deltaell}
This signal is based on an effect of neutrinos on the CMB spectrum which is 
of completely different nature from the previous one.  In the standard 
cosmology, neutrinos are assumed to be non-interacting and relativistic when
$T\approx\ev$.  This assumption has a significant consequence as follows.  
Suppose that we know what the CMB spectrum looks like 
{\it without} taking into account the free-streaming of neutrinos, and then 
consider how this spectrum gets modified by putting the free-streaming back.
While the photon-baryon plasma undergoes acoustic oscillations in the 
gravitational potential created by dark matter, the neutrinos just free-stream
with the speed of light from over-dense regions to under-dense regions.  The 
difference between the speed of this neutrino flow and the speed of 
sound-waves in the plasma causes extra phase shift in the acoustic 
oscillation.  Bashinsky and Seljak \cite{Bashinsky:2003tk} has derived an 
analytic expression for how the neutrino free-streaming affects the locations 
of CMB peaks.  A striking feature of their result is that the neutrino 
free-streaming shifts the locations of {\it all} the peaks by an {\it equal} 
amount!  For our purpose, their result can be cast conveniently as 
\cite{Chacko:2003dt}:
\beq
   \Delta \ell = -57 \left( \frac{0.23 N_\nu^{FS}}{1+0.23 N_\nu} \right)
                     \left( \frac{\delta \ell_{\rm peak}}{300} \right)
\label{eq:deltaell} ~,
\eeq
where $N_\nu^{FS}$ is the (effective) number of neutrino species which are
free-streaming at $T\approx\ev$, while $N_\nu$ is the grand total (effective) 
number of neutrino species during the same era.  $\delta \ell_{\rm peak}$ is 
the spacing between two successive peaks, which is $\approx 300$.  Note that 
the amount of shift is {\it independent} of which peak we are looking at, 
although one should keep in mind that for the first few peaks it is difficult 
to isolate this effect from larger peak shifts due to other causes.  For 
higher multipole $\ell$, this signal is very clean and unique. 

We have already explained in section \ref{subsec:deltaNnu} how $N_\nu$ 
can be different from 3 in our 
scenario.  What about $N_\nu^{FS}$?  How can it be different from $N_\nu$? 
One possibility is that $N_\nu^{FS}$ is less than $N_\nu$ due to scattering of
neutrinos with $\phi$'s.  If the scattering occurs sufficiently frequently, 
it will prevent the neutrinos from free-streaming.%
\footnote{One may think that the possibility of such scattering is disfavored
by the analyses in Refs~\cite{Hannestad:2004qu, Trotta:2004ty}.  However, 
the crucial difference between 
their model and ours is the occurrence of neutrino recoupling.  Because of
recoupling, neutrinos and $\phi$'s {\it share} energy that was originally 
carried by neutrinos, while in their model neutrinos and $\phi$'s are 
already thermalized with the rest of the universe prior to the decoupling of 
neutrinos.}
In flavor symmetry 
breaking scenarios in Ref.~\cite{Chacko:2003dt}, this signal is useful for 
probing the neutrino mass spectrum because $g$ is proportional to $m_\nu$ so 
that a heavier neutrino is less likely to free-stream.  

In our case, however, there is an additional process which must be considered.
Since our $\nu_R$ is a composite object, they can scatter with each other 
without involving extra particles like $\phi$ or any other long-range forces.
In other words, at energies below the compositeness scale $\Lambda$, the 
Lagrangian contains the operator $\bar{\nu}_R\bar{\nu}_R\nu_R\nu_R$ with a 
large coefficient, thanks to the strong nu-color interaction.  This operator 
leads to important consequences which have no analogs in the flavor symmetry 
breaking scenarios.

\section{Signal Regions In Parameter Space}
\label{sec:region}
There are two completely different cases in our scenario which must be 
separately considered.  In both cases, the nu-color sector is designed such 
that confinement produces three massless ``baryons'' which we identify as 
three right-handed neutrinos, and their masslessness is guaranteed by low 
energy chiral symmetry which survives through confinement.  Now, the two 
cases are distinguished by comparing the high energy and low energy flavor 
symmetries.  In the first case, the low energy symmetry is smaller than the 
high energy symmetry---that is, the nu-color sector undergoes chiral symmetry 
breaking, producing massless ``pions''.  We take into account a more general 
possibility that these ``pions'' may be pseudo-Goldstone bosons, treating 
their masses to be free-parameters.  Therefore, the low energy spectrum of 
a theory of this type contains these naturally light scalars as well as the 
three massless right-handed neutrinos. 

In the second case, confinement occurs without any chiral symmetry breaking, 
\ie, the original flavor symmetry of the high energy theory is completely 
preserved even after confinement.  Therefore, in this case, there are no 
light particles beneath $\Lambda$ other than the three massless right-handed 
neutrinos.

\subsection{Case With Chiral Symmetry Breaking}
\label{subsec:chiral}
In this case, we have naturally light scalars, ``pions'', which have 
interactions with neutrinos at low energies.  These interactions may be
able to induce neutrino-recoupling.  One may suspect that the results in this 
section should be identical to those of the Dirac neutrino case discussed in 
\cite{Chacko:2003dt} where they have pseudo-Goldstone bosons from flavor 
symmetry breaking instead of the ``pions''---the only difference seems our 
flavor symmetry is broken by strong dynamics instead of by the VEV of a scalar 
field.  However, that is not the only difference---as we mentioned earlier, 
the scattering property of the neutrino fluid can be very different because 
of the compositeness of $\nu_R$.  So we will focus on this point in this 
section.

After electroweak symmetry breaking and nu-color confinement, the leading 
interaction is contained in
\beq
   {\cal L}_{int} \approx v \left( \frac{\Lambda}{M} \right)^{3(n-1)/2} 
                               \nu_L \nu_R \, \e^{i\pi/f} 
                                 +\hc  ~,
\label{eq:Lint-1}
\eeq
where $\pi$ collectively denotes the ``pions'', and $v$ is the electroweak 
VEV, and $f \approx\Lambda/4\pi$.  
As we discussed in Sec.\ref{sec:fatnu}, the ratio $(\Lambda/M)^{3(n-1)/2}$ is 
fixed  to be $3\times 10^{-13}$ to get the right neutrino mass.  Therefore, 
(\ref{eq:Lint-1}) can be rewritten more nicely as 
\beq
   {\cal L}_{int} \approx g \, \nu_L \nu_R \, f \, \e^{i\pi/f} 
                                    +\hc   ~,
\eeq
where
\beq
   g \equiv \frac{m_{\nu_3}}{f}  ~,
\label{eq:g-value}
\eeq
with $m_{\nu_3} \approx 0.05\>\ev$. 

First, consider a process, $\nu_L \, \nu_R \to \pi \, \pi$.  One might 
think that this process is impossible because initially there should not be 
any $\nu_R$'s available in the universe because of the BBN constraint.  
However, it is very hard to imagine that the number density of $\nu_R$'s is 
{\it absolutely} zero; after all, the post-inflation reheating may well 
create ``quarks'' and ``gluons'' of the nu-color sector as well.  So it is 
more natural to expect that there are {\it some} amount of $\nu_R$'s.  In fact, 
the BBN bound allows $N_\nu = 3.2$ at the $2\sigma$ level, so as long as 
the energy density of the $\nu_R$'s is less than a several percent of that of 
the $\nu_L$'s, there is no conflict with BBN.  There are several ways by which 
the $\nu_R$ energy density becomes suppressed.  For example, since the nu-color
sector is neutral under the standard-model gauge group, the latent heat from 
the electroweak and/or QCD phase transitions, if any, is transfered only to 
the standard-model sector.  In any case, to account for this possible $\nu_R$ 
suppression, we introduce $r \equiv n_{\nu_R}/n_{\nu_L}$.  The BBN constraint 
then tells us that if 
the temperatures of the $\nu_L$ and the $\nu_R$ are the same, $r$ should be 
less than 0.1 or so, but actually $r$ can even be $\order{1}$ if the $\nu_R$ 
temperature is less than about a half of the $\nu_L$ temperature, which is 
perfectly fine because the two sectors are not in thermal contact before 
neutrino-recoupling.  So, $r$ may be $\order{1}$, 
$\order{0.1}$, or smaller, but it begins to look very unnatural as $r$
becomes too small.           

Now, in order for the process, $\nu_L \, \nu_R \to \pi \, \pi$, to 
induce neutrino-recoupling, the pions must be relativistic, since otherwise 
the cross-section would be highly suppressed by a tiny phase space.  
However, even in the relativistic limit, the cross-section for this process 
is {\it not} given by $\sim g^4/T^2$, due to the derivatively-coupled nature 
of $\pi$; it actually goes as $g^4/m_{\nu_3}^2$.  Thus, the rate is given by
\beq
   \Gamma_{\nu_L\nu_R \to \pi\pi} 
           \approx \frac{rg^4 T^3}{16\pi m_{\nu_3}^2}  ~.
\label{eq:nunu-pipi}
\eeq
Unfortunately, this rate dies faster than the expansion rate as $T$ drops, 
so this process cannot recouple neutrinos.  Instead, this process is more 
important at higher temperatures.  Of course, it does not become 
indefinitely more important as $T$ increases, because the above formula 
assumes $T<\Lambda$.  The easiest way to understand the qualitative behavior 
of the cross-section for $T> \Lambda$ is 
to look at the similar process in a theory where chiral symmetry 
is spontaneously broken by a VEV of a strongly-self-coupled scalar.  In that
case, the ``Higgs'' mode with mass of $\order{4\pi\times\text{VEV}}$ comes in 
to the process, and as a result the behavior of the cross-section changes 
from $\propto 1/m_\nu^2$ to $\propto 1/T^2$ for energies larger than 
$\order{4\pi\times\text{VEV}}$.  Similarly, in our case, some resonances 
come in to the process at $E \approx 4\pi f \approx \Lambda$ and changes the 
the cross-section from $\propto 1/m_\nu^2$ to $\propto 1/T^2$.  Therefore, 
the rate is actually maximal at $T\approx\Lambda$, compared to the 
expansion rate.

Similarly, consider another process, $\nu_L \, \nu_R \to \pi \, \pi \, \pi$.  
The rate of this process is given by
\beq
   \Gamma_{\nu_L\nu_R \to \pi\pi\pi} 
           \approx \frac{T^2}{16\pi^2 f^2} 
                      \Gamma_{\nu_L\nu_R \to \pi\pi}   ~.
\label{eq:nunu-pipipi}
\eeq 
where the $16\pi^2$ accounts for the 3-body phase space.  Therefore, although 
this process is subdominant to the previous process at low temperatures, it 
becomes as important when $T$ reaches near $4\pi f \approx \Lambda$. This is 
in fact expected because the theory becomes strongly-coupled at this scale. 

Now, if these two processes are occurring at temperatures near $\Lambda$, 
then all of $\nu_L$, $\nu_R$ and $\pi$ are brought into equilibrium without 
chemical potentials.  This will lead to a value of $N_\nu$ that is too large
at the BBN era, so we must avoid it.  Hence, we evaluate the rate 
(\ref{eq:nunu-pipi}) at 
$T\approx\Lambda$, and demand that it be smaller than the expansion rate:
\beq
   \frac{rg^4\Lambda^3}{16\pi \, m_{\nu_3}^2} \lsim \frac{\Lambda^2}{M_{pl}}
     ~~~\Longrightarrow~~~
   \Lambda \gsim \left( 16\pi^3 r \, m_{\nu_3}^2 M_{pl} \right)^\frac13  
           \approx r^\frac13 \> \gev
\label{eq:BBNbound}  ~,
\eeq
where we have used (\ref{eq:g-value}). 

On the other hand, in order for the $\nu_R$ compositeness to play any role 
in modifying the CMB spectrum, the rate of the $\nu_R\nu_R \to \nu_R\nu_R$ 
must be appreciable at $T\approx\ev$, because this is the leading process that 
can probe the compositeness.  At energies beneath $\Lambda$, the 
compositeness of the $\nu_R$ yields the operator
\beq
   {\cal L}_{int} \approx \left( \frac{4\pi}{\Lambda} \right)^2 
                             \bar{\nu}_R\bar{\nu}_R \nu_R\nu_R   ~,
\label{eq:4nu_Rop}
\eeq     
where we have estimated the coefficient using Naive Dimensional Analysis (NDA)
\cite{Weinberg:1978kz, Manohar:1983md, Georgi:kw, Georgi:1986kr}.  Thus, 
the rate of the $\nu_R \nu_R$ scattering is given by
\beq 
   \Gamma_{\nu_R\nu_R\to\nu_R\nu_R} 
         \approx \frac{(4\pi)^4T^5}{16\pi\Lambda^4}   ~,
\eeq
and demanding that this rate be larger than the expansion rate at 
$T \approx \ev$ gives rise to
\beq
   \Lambda < 4\pi \left( \frac{\ev^3 M_{pl}}{16\pi} \right)^\frac14
             \approx 10\> \mev   ~.
\eeq     
Note that this becomes consistent with the BBN bound (\ref{eq:BBNbound}) 
{\it only} if $r < 10^{-6}$!  Unless there is a good reason why $r$ can be
made so tiny, this is quite implausible.         

Therefore, we conclude that if the underlying nu-color sector has chiral 
symmetry breaking, BBN requires the confinement scale $\Lambda$ to be very 
high as in (\ref{eq:BBNbound}), and in fact it is so high that no sign of the 
compositeness can be seen at low energy scales relevant for the
CMB (\ie, $T \approx \ev$), and consequently the CMB signals are 
indistinguishable from those in the Dirac neutrino case analyzed in 
Ref.~\cite{Chacko:2003dt} where small neutrino masses arise from 
low-energy spontaneous flavor-symmetry breaking by the VEV of a 
(weakly-coupled) scalar.  The reader should look at Ref.~\cite{Chacko:2003dt} 
for details, but in short the signals in their case arise from the process, 
$\nu_L\nu_R \to G$, where $G$ is a pseudo-Goldstone boson from the 
flavor-symmetry breaking.  This occurs with the rate 
\beq 
   \Gamma_{\nu_L\nu_R \to G} \approx \frac{r g^2 m_G^2}{16\pi T} ~,  
\eeq
where $m_G$ is the mass of the $G$.  
This reaction clearly can induce recoupling.    

It is unfortunate that we cannot distinguish these two different underlying 
mechanisms for generating small neutrino masses, but a positive way of 
viewing this is that a 
nu-color sector with chiral symmetry breaking offers a very natural, simple 
way to stabilize the very low symmetry-breaking scale appearing in such 
theories, alternative to the supersymmetric stabilization discussed in 
Ref.~\cite{Chacko:2003dt}.

\subsection{Case Without Chiral Symmetry Breaking}
\label{subsec:nochiral}
In this case, the three $\nu_R$'s are the only massless particles, and all 
other ``hadrons'' have masses of $\order{\Lambda}$.  This means that if 
$\Lambda > \mev$, those particles are too heavy to participate in 
neutrino-recoupling, 
because recoupling must occur at a temperature {\it below} $\order{\mev}$ 
to avoid having too many thermalized relativistic degrees of freedom during 
BBN.  One may wonder if a 4-fermion operator, 
$\bar{\ell}\bar{\nu}_R \ell\nu_R$, can induce neutrino recoupling.  The answer 
is no, because the natural size of the coefficient of this operator is 
$\order{10^{-25}}$ because of the compositeness of the $\nu_R$, and this is 
too small to do anything. 

What about the operator (\ref{eq:4nu_Rop})?  Can it induce recoupling?  
Let us consider a process, 
$\nu_L\,\nu_L \to \bar{\nu}_R \, \bar{\nu}_R$, where we put two mass-insertions
in the initial state to flip the helicity of each $\nu_L$.  Note 
that processes with only {\it one} mass insertion such as 
$\nu_L\,\nu_R \to \nu_R \, \bar{\nu}_R$ are forbidden by angular momentum 
conservation.  Hence, the process, 
$\nu_L\,\nu_L \to \bar{\nu}_R \, \bar{\nu}_R$, (and the ``bar'' of this 
process) is {\it the} leading process that 
can couple $\nu_L$ and $\nu_R$ via the operator (\ref{eq:4nu_Rop}).  
The rate of this process is given by 
\beq
   \Gamma_{\nu_L\nu_L\to\nu_R\nu_R} 
      \approx \frac{(4\pi)^4}{16\pi \Lambda^4} \frac{m_\nu^4}{T^4} T^5   ~.
\eeq
Since this rate dies slower than the expansion rate, it can induce 
recoupling.  Comparing this with $T^2/M_{pl}$, we obtain the recoupling 
temperature:
\beq
   T_{rec}^{\nu_L\nu_L\to\nu_R\nu_R} 
         \approx \frac{16\pi^3 M_{pl} \, m_\nu^4}{\Lambda^4}   
          \approx 10 \left( \frac{\mev}{\Lambda} \right)^4
                       \left( \frac{m_\nu}{0.05\> \ev} \right)^4 \> \ev  ~.
\eeq
BBN requires $T_{rec} < \mev$, leading to
\beq
    \frac{\Lambda}{100\> \kev} \gsim \frac{m_\nu}{0.05\> \ev}  ~.
\label{eq:4-fermi-BBN}
\eeq
If the neutrino masses are not degenerate, this bound must be applied for the 
heaviest one. On the other hand, generating a CMB signal needs 
recoupling to occur for $T>\ev$, leading 
to
\beq
   \frac{\Lambda}{\mev} \lsim \frac{m_\nu}{0.05\> \ev} ~.
\label{eq:4-fermi-rec}
\eeq
Here, we should recall that the $0.05\>\ev$ is the mass of the heaviest 
neutrino if the mass spectrum is hierarchical or inverse hierarchical.  In 
these cases, the recoupling condition (\ref{eq:4-fermi-rec}) implies 
$\Lambda < \mev$, and in particular {\it this invalidates the BBN bound 
(\ref{eq:4-fermi-BBN})} because BBN occurs at $T\approx\mev$ which is now 
higher than $\Lambda$, so the use of the operator (\ref{eq:4nu_Rop}) is not 
justified.  On the other hand, if the mass spectrum is degenerate, 
$m_\nu$ can be larger than $0.05~\ev$, so (\ref{eq:4-fermi-rec}) may allow    
$\Lambda > \mev$ in which case the bound (\ref{eq:4-fermi-BBN}) applies.   
However, even in the the degenerate case, $m_\nu$ is not expected to be much 
larger than $0.5\>\ev$.%
\footnote{A limit on the sum of neutrino masses from 
large-scale structure is about $0.5$-$2$ eV 
\cite{Spergel:2003cb, Tegmark:2003ud, Hannestad:2003ye}.
However, all of these bounds assume that neutrinos free-stream, which is not 
necessarily true in our scenario.  A more direct bound, $m_\nu < 2.2\>\ev$, 
has been obtained from tritium $\beta$-decay experiments 
\cite{Weinheimer:tn, Lobashev:tp, Bonn:tw}.  
Note that this bound should apply to {\it each} of the three mass
eigenstates, since neutrino oscillation data imply 
$\delta m^2 \ll \ev^2$ and large mixing angles\cite{Beacom:2004yd}.
}
In other words, $\Lambda$ is not expected in any case to be larger than
a several $\mev$.%
\footnote{
Such a low value of $\Lambda$ and 
correspondingly a low value of $M$ might have constraints and signals in 
collider experiments.  We will discuss those in some detail in 
section \ref{sec:collider}.  In this section we will focus on constraints and 
signals in cosmological and astrophysical circumstances.       
}  

So, let us take $\Lambda < \mev$ and see where it leads to.  It seems, then,  
that there are other routes for neutrino-recoupling and danger of messing up 
BBN, because now ``hadrons'' from the nu-color confinement are 
light enough that they may participate in recoupling and BBN.  Let us 
examine this point.  First, if $T<\Lambda$, then we can integrate out all the 
nu-color composites except the $\nu_R$'s which are massless; we have already 
analyzed this case above.  Therefore, consider a case, $\Lambda< T < M$.  In 
this range, recoupling should occur via the following interaction that is 
obtained from (\ref{eq:op-at-M}) with $\vev{h}=v$: 
\beq
   {\cal L}_{int} \approx \frac{v}{M^{3(n-1)/2}}\nu_L \> qq\cdots q   ~,
\eeq
where the $q$'s here are now unconfined ``quarks''.  However, even for the 
minimal case of $n=3$, this is a highly irrelevant operator and the rate for a 
process like $\nu_L \bar{\nu}_L \to qq\bar{q}\bar{q}$, for instance, 
dies away too quickly, so neutrinos cannot recouple.  Of course it is even 
worse for larger $n$.  Therefore, we conclude that the recoupling temperature 
should be higher than $M$.  It should be emphasized here that if a process 
involves an irrelevant coupling, it 
cannot lead to recoupling, because it requires too many factors of $T$ 
in the numerator to compensate a power of the mass scale appearing 
in the denominator.  The 
only way to kill a large power of $T$ in the numerator is to put mass 
insertions, $m_\nu^2/T^2$, but for $T>\Lambda$ there exist no such mass 
terms.  Therefore, in order to have recoupling, we need  
a ``UV theory'' which generates 
(\ref{eq:op-at-M}) without involving any irrelevant operators.  

It turns out, however, that we can extract quite a bit of information 
without knowing exactly what the UV theory is.  
First of all, the UV theory should involve not only $q$'s but also some 
additional, ``messenger'' fields, collectively called $X$, which 
communicate the standard-model sector with the nu-color sector; interactions 
of $X$ with $q$ and $\nu_L$ should be designed such that at energies beneath 
the $X$ mass, integrating out $X$ will yield the operator (\ref{eq:op-at-M}) 
(and most likely many other operators of much higher dimension).  Secondly, 
as we argued above, interactions among $X$, $q$ and $\nu_L$ cannot involve 
irrelevant operators in order to have recoupling.  
For simplicity, we assume that all relevant operators 
such as the $X$ mass just involve a single mass scale $m_X$.  

Now, consider a diagram where we have one $\nu_L$ and $n$ $q$'s in the 
external states while the internal state involves only $X$'s.  This becomes the 
operator (\ref{eq:op-at-M}) at energies below $m_X$.  Here we see that the 
scale $M$ in (\ref{eq:op-at-M}) should be identified as
\beq
   \frac{v}{M^{3(n-1)/2}} \approx \frac{\lambda}{m_X^{(3n-5)/2}}    ~,
\label{eq:M-m_X}
\eeq
where $\lambda$ is the product of all the coupling constants 
appearing in the diagram.  Also, the relation (\ref{eq:M-Lambda}) allows us to 
rewrite this in terms of $\Lambda$ instead of $M$:
\beq
   \frac{m_{\nu_3}}{\Lambda^{3(n-1)/2}} 
                    \approx \frac{\lambda}{m_X^{(3n-5)/2}}    ~.
\label{eq:Lambda-m_X}
\eeq  
One thing we have to worry about here is whether or not the addition of $X$ 
messes up the confinement dynamics.  To answer this question, note that 
in order to get neutrino masses, the operator (\ref{eq:op-at-M}) should be 
{\it already} there when the ``$qq\cdots q$'' confines into $\nu_R$.  
This order cannot be reversed.  Therefore, we must impose
\beq
   m_X \gg \Lambda
\label{eq:m_X-Lambda-1} ~,
\eeq
so that we can integrate out $X$ to get (\ref{eq:op-at-M}) before we reach 
the confinement scale $\Lambda$.  This also guarantees that the confinement 
dynamics is not affected by the addition of $X$.  Combining 
(\ref{eq:Lambda-m_X}) and (\ref{eq:m_X-Lambda-1}), we obtain a very simple 
bound:
\beq
   \lambda \gg \frac{m_{\nu_3}}{\Lambda} 
     ~~~\Longrightarrow~~~  
       \frac{\lambda}{10^{-7}} \gg \frac{\mev}{\Lambda}  ~. 
\label{eq:m_X-Lambda-2}
\eeq

Now, note that this same diagram can also cause the process, 
$q\,\nu_L \to (n-1)\bar{q}$, to occur.
Once this happens, then 
various reactions like  $qq \leftrightarrow qq$, $qq \leftrightarrow XX$, 
etc. start occurring very rapidly and everything becomes thermalized without 
chemical potentials, which is bad for BBN.  This second step occurs really 
quickly because these reactions can be mediated by exchanging nu-color gluons, 
leading to the rate of $\order{T/16\pi}$ or larger, which is much faster than 
the rate of the initial triggering process, $q\,\nu_L \to (n-1)\bar{q}$.  Of 
course, we also need $T/16\pi \gg \Lambda$ so that 
that they do not confine before thermalization occurs, but since we are 
analyzing here the case where $T>m_X \gg \Lambda$, $T/16\pi$ is bigger than 
$\Lambda$ by assumption.  Therefore, we have to make sure that the initial 
triggering reaction, $q\,\nu_L \to (n-1)\bar{q}$, does not happen before BBN.  
If none of the virtual $X$'s in this process is a vector or a 
derivatively-coupled scalar (otherwise recoupling cannot occur), then the rate 
is simply given by dimensional analysis as 
\beq
   \Gamma_{q\nu_L\to (n-1)q} 
         \approx \frac{r \lambda^2 T}{16\pi (16\pi^2)^{n-3}}  ~,
\label{eq:qnutoqqqq}
\eeq
where the factor of $(16\pi^2)^{n-3}$ accounts for the ($n-1$)-body 
phase space, while $r$ takes into account the effective fraction of the
number density of $q$'s with respect to that of $\nu_L$, analogous to the $r$ 
in section \ref{subsec:chiral}.  Demanding this rate to
be less than the expansion rate at $T\approx \mev$ gives
\beq
   \lambda \lsim r^{-1/2}(4\pi)^{n-2} \times 10^{-11}   ~.
\eeq   
Note that this is inconsistent with (\ref{eq:m_X-Lambda-2}) and 
$\Lambda<\mev$ {\it unless}
\beq
   r \ll (4\pi)^{2n-4} \times 10^{-8} ~.
\eeq
For $n=3,~5$ this restricts $r$ as $r\ll 10^{-6},~10^{-2}$, respectively, 
while $n=7$ gives no restriction.     

However, if we suppress the $q\,\nu_L \to (n-1)\bar{q}$ process by, say, 
taking $n=7$, then there will be other processes dominating 
the physics.  For example, consider a process, $q\,\nu_L \to X$.  Since the 
``UV theory'' necessarily includes a term like $\nu_L q X$ in order for the 
standard-model and nu-color sectors to be connected via non-irrelevant 
interactions, this process inevitably exists.  Then, the rate of this process 
is given by
\beq
   \Gamma_{q\nu_L  \to X} 
       \approx \frac{r \lambda'^2 \tilde{m}_X}{16\pi} \frac{\tilde{m}_X}{T} ~,
\label{eq:qnutoX}
\eeq
where $\lambda'$ is the product of all the coupling constants
appearing in the amplitude, while $\tilde{m}_X$ is the effective 
``temperature-corrected'' mass of the $X$, \ie, 
$\tilde{m}_X^2 = m_X^2 + a^2 T^2$ with some constant $a$.  Since 
$X$ is strongly interacting via the nu-color interaction, $a$ is expected to 
be $\order{1}$.  So, remembering the assumption $T > m_X$, we 
approximate $\tilde{m_X}$ as $\tilde{m}_X \approx aT$. 
Then, the process $q\nu_L\to X$ will lead to recoupling at the temperature
\beq
   T_{rec}^{q \nu_L  \to X} 
      \approx  \frac{r \lambda'^2 a^2 M_{pl}}{16\pi}  ~.
\eeq  
Once some amount of $X$'s is produced from this process, various reactions 
like $q X \leftrightarrow q X$, $X \leftrightarrow qq$, etc. will 
begin rapidly via the nu-color gauge interaction and immediately thermalize 
all of $\nu_L$, $q$ and $X$s with no chemical potentials. Therefore, we 
have to make sure that the initial triggering process, $q\nu_L\to X$, 
occurs after BBN.  Demanding that the recoupling temperature is less than 
$\mev$, we obtain
\beq
   r \lambda'^2 \lsim 10^{-19}  ~,
\eeq
where we have ignored $a$ since it is $\order{1}$.
Since our very motivation to consider composite neutrinos is to avoid the 
very tiny Yukawa couplings for the Dirac neutrinos, we do not want to 
reintroduce a small Yukawa coupling here.  So, if we assume that $\lambda'$
is no smaller than the electron Yukawa coupling, $r$ must be smaller than 
$10^{-7}$!  Unless there is a good reason why $r$ could be made so small, 
this is quite implausible.  

Therefore, we conclude that in any reasonable scenario, BBN does not allow 
processes involving $X$ or $q$ to induce recoupling.  This means that
the BBN constraint forces (some of) the various assumptions we made which 
led to (\ref{eq:qnutoqqqq}) and (\ref{eq:qnutoX}) to be violated.  Note that 
we made those assumptions so that those processes could induce recoupling.  
Thus, our analysis shows that if recoupling occurred for $T> \Lambda$, it would 
be incompatible with BBN.%
\footnote{Depending on how those assumptions are 
violated, the actual bounds from BBN vary.  Here, we simply assume that a 
conflict with BBN is avoided by adjusting parameters in the ``UV theory'' 
that generates the operator (\ref{eq:op-at-M}).  But, having giving up 
recoupling, this can be now done easily, because the tension was always 
between BBN and recoupling.
}  
Therefore, recoupling can occur only for $T<\Lambda$ 
and the only reaction that can recouple neutrinos is the 
$\nu_L \nu_L \to \bar{\nu}_R \bar{\nu}_R$ process (and the ``bar'' of this 
process), 
which we have discussed at the beginning of this section.  

What kind of CMB signals do we predict?  
First, a $\Delta N_\nu$ signal will not exist.  Since $T_{rec} < \Lambda$, 
there are no heavy particles which are in equilibrium with neutrinos after 
recoupling.  The confinement phase transition may produce some latent heat, 
but that occurs at $T \approx \Lambda$ which is before recoupling, and 
therefore has no consequence.  This can be a very useful prediction, 
because {\it if} we see a $\Delta N_\nu$ signal, then we know that the 
theory---if it is a composite neutrino theory---{\it must} have chiral 
symmetry breaking.  

In contrast, a $\Delta \ell$ signal should be very robust.  As long as 
the bound (\ref{eq:4-fermi-rec}) is satisfied for a mass eigenstate, we 
{\it always} have a $\Delta\ell$ signal from the scattering of any two 
neutrinos of {\it that} eigenstate, because, for instance, the rate of 
the process, $\nu_R\nu_R \to \nu_R\nu_R$, is larger by $T^4/m_\nu^4$ than 
the rate of the $\nu_L\nu_L \to \bar{\nu}_R\bar{\nu}_R$, which is already 
ensured to be faster than the expansion rate by (\ref{eq:4-fermi-rec}).    
Furthermore,
since the bound (\ref{eq:4-fermi-rec}) depends on $m_\nu$, the $\Delta \ell$ 
signal can probe the neutrino mass spectrum very sensitively!  For example, 
only in the case of the hierarchical spectrum does exists the region where 
we observe a $\Delta\ell$ signal corresponding to only {\it one} neutrino 
species scattering, \ie, $\Delta\ell \approx 8$.  Similarly, if the 
$\Delta\ell$ signal indicates that two 
neutrino species are scattering ($\Delta\ell\approx 16$), then we will know 
that the spectrum cannot be degenerate. 
If we see three neutrinos scattering ($\Delta\ell\approx 24$), then we will 
have no clue.

\section{Constraints From Terrestrial Experiments}
\label{sec:collider}
In section \ref{subsec:nochiral}, we found that in theories without chiral 
symmetry breaking, CMB signals are present only for a very low confinement 
scale $\Lambda$ and correspondingly a low messenger mass scale $m_X$.  
So, it is important to check 
if there are constraints from terrestrial experiments.

To discuss terrestrial bounds, we need to specify how the standard model 
couples to the nu-color sector, because we can certainly go beyond the scale 
$M$, so the operator (\ref{eq:op-at-M}) must be ``resolved'' in terms of 
renormalizable operators.  Recall the requirement that the 
communication should not involve any irrelevant operator.  Interestingly and
fortunately, everything we need for analyzing physics in terrestrial 
experiments is uniquely fixed.  Schematically, it is
\beq
   {\cal L} = \alpha \, \ell N h  + m_N \, N N^c 
                         + (\text{Terms with~} N^c, q, X)   
\label{eq:messengerlag} ~,
\eeq
where we have introduced two neutral fermions, $N$ and $N^c$.  Note that 
it is necessary to introduce a pair of $N$ and $N^c$ for each generation in 
order to give masses to all three generations.  Therefore, both $\alpha$ and 
$m_N$ in (\ref{eq:messengerlag}) are $3\times 3$ matrices.  However, after 
electroweak symmetry breaking, $\vev{h} = (0,~v+h^0)$, we can always first 
rotate $N$ and $N^c$ to diagonalize $m_N$, then rotate $e_L$, $\nu_L$ and $N$ 
to diagonalize the charged lepton mass matrix and $\alpha$, while 
simultaneously rotating $N^c$ in the opposite manner as $N$ such that $m_N$ 
remains diagonal.  So, we can assume without loss of generality that both 
$\alpha$ and $m_N$ are diagonal and all the flavor violations are put in 
$U_{MNS}$, which we will not write explicitly below because we know that 
matrix elements of $U_{MNS}$ are $\order{1}$.     

Therefore, for {\it each} generation, we have a $3\times 3$ mass matrix 
spanned by 
$\nu_L$, $N$ and $N^c$.  This can be ``diagonalized'' as
\beq
   \alpha v \, \nu_L N + m_N \, N N^c + \hc
         = 0\cdot \nu_0 \nu_0 + m_D \, N N' + \hc  ~,
\eeq
where the mass ``eigen''states, $\nu_0$ and $N'$, are given by
\beq
   \nu_L &=&  \nu_0 \cos\theta_N + N'\sin\theta_N  ~,\nn\\ 
   N^c   &=& -\nu_0 \sin\theta_N + N'\cos\theta_N  ~,   
\label{eq:mixing}
\eeq
and the mass, $m_D$, and the mixing angle, $\theta_N$, by
\beq  
   m_D = \sqrt{\alpha^2 v^2 +m_N^2}   ~~~,~~~
   \sin\theta_N = \frac{\alpha v}{m_D}  ~~~,~~~
   \cos\theta_N = \frac{m_N}{m_D}   ~.
\eeq
Therefore, at this point, each generation consists of a massless Weyl fermion 
$\nu_0$ and a massive Dirac fermion $\Psi_N$ made of $N$ and $N'$.  We will 
not consider a case where $m_D \lsim \mev$, because the very motivation for 
considering composite neutrinos is to explain why the Dirac masses of 
neutrinos are so small compared to those of the charged fermions.  Hence, we 
assume that $m_D$ is no smaller than the electron mass.

Now, recall that in the standard cosmology, neutrinos decouple from the rest 
of the universe at $T\approx\mev$.  In our case, because of the mixing 
(\ref{eq:mixing}), $\nu_0$ decouples at 
$T\approx \mev/(\cos\theta_N)^{4/3}$, strictly speaking.  However, since 
$\cos\theta_N \simeq 1$ as we will see shortly, we just say $\nu_0$  
decouples at the standard temperature, $T\approx \mev$.  
The situation with $N'$ is more complicated.  First of all, if they decouple 
while they are relativistic, then the the decoupling temperature is 
given by $T \approx \mev/(\sin\theta_N)^{2/3}$.  However, since 
$m_D \gsim \mev$, they soon become non-relativistic before BBN.  
Alternatively, they may be already non-relativistic when they decouple.  In 
either case, a $N'$ decays rapidly into 
$\nu_0$'s via $N' \to \nu_0 \nu_0 \bar{\nu}_0$ with the lifetime of order 
$(m_\mu/m_D)^5/\sin^2\theta_N$ times the muon lifetime, 
$10^{-6}$ sec.  In order not to screw up BBN, they must decay {\it before} 
$\nu_0$'s decouple, so we impose
\beq
   \frac{m_\mu^5}{m_D^5\sin^2\theta_N} \times 10^{-6}
    ~\text{sec}
   \lsim 10^{-2} ~\text{sec}  ~,  
\eeq
which gives
\beq
   m_D \sin^\frac25\theta_N \gsim 10\>\mev   ~.  
\label{eq:m_Dbound}
\eeq 
  
\begin{itemize}
 \item{Muon decay:\\
  First, note that the mixing angle clearly cannot be too large, because 
  if $\nu_L$ is made predominantly of $N'$, then 
  (\ref{eq:m_Dbound}) would be in conflict with the direct upper bound,  
  $m_\nu < 2.2 \> \ev$, obtained by combining tritium $\beta$-decay 
  experiments with neutrino oscillation data.  We will sharpen this point 
  below. 
   
  The most precise measurement of $G_F$ comes from muon decay.  In our case, 
  if $m_D < m_\mu$, there are three decay 
  modes:~$\mu \to e \nu_0 \bar{\nu}_0$,  
  $\mu \to e \nu_0 \bar{N'}$ and $\mu \to e N' \bar{\nu}_0$, with the rates 
  proportional to $\cos^4\theta_N$, $\sin^2\theta_N$, $\sin^2\theta_N$, 
  respectively.  
  We neglect the case where both of the final neutrinos are $N'$, because 
  $\theta_N$ should be small as we argued above.  When it decays to a $N'$, 
  the rate depends on $m_D$ rather than $m_e$, because $m_D >\mev> m_e$.  
  Therefore, for $m_D < m_\mu$, the muon decay rate is given by 
  $G_F^2 m_\mu^5/192\pi^3$ times  
  \beq
     \cos^4\theta_N \left( 1 - \frac{8 m_e^2}{m_\mu^2} 
                              + \order{\frac{m_e^4}{m_\mu^4}} \right) 
       + 2\sin^2\theta_N \left( 1 - \frac{8 m_D^2}{m_\mu^2} 
                               + \order{\frac{m_D^4}{m_\mu^4}} \right)  \nn\\ 
  \eeq
  Since $G_F$ is currently measured down to 4 decimal 
  places\cite{Hagiwara:fs}, we must impose $\sin\theta_N < 10^{-2}$.    
  However, combining this with (\ref{eq:m_Dbound}) requires $m_D$ to be at
  least $100\>\mev$, so this analysis which has assumed $m_D < m_\mu$ is not 
  valid.  Therefore, it must be that $m_D > m_\mu$ and a muon cannot decay 
  into $N'$.  Then, the standard muon decay rate simply gets multiplied 
  by $\cos^4\theta_N$, and requiring that $1-\cos^4\theta_N < 10^{-4}$  
  gives $\theta_N < 10^{-2}$.  Therefore, the bound from the muon decay is 
  summarized as 
  \beq
      m_D > m_\mu ~~~\text{and}~~~ \theta_N < 10^{-2}   ~.
  \label{eq:thetabound}
  \eeq  

  Next, since $\nu_L$ and $N^c$ mix, a neutrino from a muon $\beta$-decay can 
  further decay into $q$ and $X$ via interactions indicated by $(\cdots )$ 
  in (\ref{eq:messengerlag}).  Most of them are harmless because we can 
  choose them to be however small we want.  There is one process, however, 
  for which we may not have this freedom.   It is the process where one $\nu$ 
  goes to $n$ $q$'s, which we need to generate the operator 
  (\ref{eq:op-at-M}).  The rate of this process is given by
  \beq     
     \approx \frac{\lambda^2}{(16\pi^2)^{n-1}} \frac{G_F^2 m_\mu^5}{192\pi^3}
  \eeq
  Thus, for the worst case of $n=3$ and $\lambda =\order{1}$, 
  the suppression is $\order{10^{-4}}$, which is right at the experimental 
  bound.  So, the $G_F$ measurement from muon decay requires 
  $\lambda$ to be less than $\order{1}$ for $n=3$, and no restriction for 
  higher $n$.  This is not constraining our theory.  
 }
 \item{Invisible $Z$-width:\\ 
  The $Z$ decay rate into a $\nu_0 \bar{\nu_0}$ pair now gets multiplied by 
  $\cos^4\theta_N$.  The $Z$ can also decay into a $\nu_0$-$N'$ pair if 
  $m_D < m_Z$, but this rate is proportional to $\sin^2\theta_N$, which is 
  very small thanks to (\ref{eq:thetabound}).   Therefore, the modification 
  in the invisible $Z$-width is less than 1 part in $10^4$, which is   
  smaller than the experimental error in the direct measurement of the 
  invisible $Z$-width---actually even smaller than the error in the 
  $N_\nu$ fit by LEP\cite{Hagiwara:fs}.  So, there is no additional 
  constraint from the $Z$-width.   
 }
 \item{Lepton flavor violation in $\mu \to e\gamma$:\\
  Note that the diagrams with an internal $\nu_0$ vanishes by GIM mechanism.  
  When the internal line is $N'$, the leading diagrams which do not 
  vanish by GIM need two insertions of $m_D$ in the internal line.  This 
  means we need one insertion of the muon mass outside, since this decay 
  is a magnetic dipole transition, which flips the helicity.  Therefore, the 
  rate is given by
  \beq
     \Gamma_{\mu\to e\gamma} 
         \approx \frac{1}{16\pi} 
                    \left( \frac{e\,(g\sin\theta_N)^2}{16\pi^2}
                          \frac{m_\mu m_D^2}{m_W^4}  \right)^2 m_\mu^3  
  \eeq
  We should compare this to the standard muon decay rate:
  \beq
     \Gamma_{\mu\to e\nu\bar{\nu}} \simeq \frac{G_F^2 m_\mu^5}{192\pi^3}
              \simeq  \frac{g^4 m_\mu^5}{6144\pi^3 m_W^4}  ~.
  \eeq    
  So,
  \beq
    \frac{\Gamma_{\mu\to e\gamma}}{\Gamma_{\mu\to e\nu\bar{\nu}}}
      \approx 10^{-2} \times \left( \frac{m_D \sin\theta_N}{m_W} \right)^4
      \approx  10^{-12} \times \left( \frac{m_D\sin\theta_N}{100~\mev} 
                                 \right)^4 ~.
  \eeq  
  Comparing this with the current strictest experimental bound, 
  $10^{-12}$ \cite{Dohmen:mp, Ahmad:1988ur}, we obtain an upper bound for 
  $m_D$:
  \beq
     m_D \lsim \frac{100 ~\mev}{\sin\theta_N}  ~.
  \label{eq:m_Dupper}  
  \eeq  
  We see that there is a large allowed region in the $m_D$-$\theta_N$ space 
  where all the bounds, (\ref{eq:m_Dbound}), (\ref{eq:thetabound}) and 
  (\ref{eq:m_Dupper}), are satisfied.
 }
 \item{Events with missing (transverse) energy:\\  
  Note that any event with an $N'$ emission from a $W$ or $Z$ looks identical 
  to the similar event with a $\nu_0$ emission.  The problem is that due to 
  the bound (\ref{eq:thetabound}), the rate of the former process is always 
  less than $10^{-4}$ of that of the latter.  If $m_D$ is sufficiently large,
  the rate will be further suppressed by a small phase space.
  Therefore, it is extremely hard to isolate this type of processes. 

  Next, consider an event where a $N$ is emitted from a Higgs. 
  Note that the coupling of the Higgs to a $\nu_0$-$N$ pair is given by 
  $\alpha\cos\theta_N \simeq m_D\sin\theta_N/v$ rather than $m_D/v$.  
  Therefore, because of the bound (\ref{eq:m_Dupper}), the Higgs coupling to 
  the $\nu_0$-$N$ pair is weaker than that to a $s$-$\bar{s}$ pair, 
  so again it will be very difficult to see this kind of events.     
 }
\end{itemize}

\section{Conclusions}
\label{sec:conc}
The idea of composite right-handed neutrinos is quite interesting, and it 
provides a very natural, simple rationale for why Dirac neutrinos can be so 
light.  
While it is impossible to test this idea in laboratory experiments, it is 
possible that there are signals in the CMB spectrum if the compositeness 
scale is sufficiently low.            

In scenarios where the underlying strong dynamics breaks chiral symmetry, 
we have found that the interactions between the neutrinos and the ``pions'' 
from the chiral symmetry breaking are quite significant so that the 
constraint from BBN requires the compositeness scale to be 
very high ($\gsim \gev$) compared to the relevant energy scale for the CMB 
physics ($T \approx \ev$).  Consequently, there are no signs of the neutrino
compositeness which can be seen in the CMB---although this scenario does
have signals the CMB, they are indistinguishable from the signals in scenarios 
discussed in Ref.~\cite{Chacko:2003dt} where the small Dirac neutrino masses 
arise from spontaneous breaking of low-energy flavor symmetry.  Therefore, 
unfortunately, the CMB signals are not useful to probe the neutrino 
compositeness.  However, it is still theoretically quite interesting in the 
sense that a composite $\nu_R$ theory with chiral symmetry breaking can offer 
an ideal mechanism to stabilize the low VEV present in those flavor breaking 
scenarios.   

On the other hand, if the underlying strong dynamics does not undergo 
chiral symmetry breaking, a very robust CMB signal can arise---a large 
$\Delta\ell$ signal can be observed in the {\it entire} allowed region in the 
parameter space thanks to the large scattering cross-section among composite 
objects.  And the allowed region is very large.  This is a great virtue, 
because the $\Delta\ell$ signal is a very 
unique and clean signal.  We have also found that the $\Delta\ell$ signal is 
very sensitive to the mass spectrum of neutrinos, providing the exciting 
possibility of determining the neutrino mass spectrum from CMB observations.  

In contrast, a $\Delta N_\nu$ signal is predicted to be absent.  This is 
also a striking prediction in the sense that if we observe a $\Delta N_\nu$ 
signal, it will mean that the underlying theory---if it is a composite 
theory---{\it must} have chiral symmetry breaking.  In this view, the 
difference between the two cases---with or without chiral 
symmetry breaking---is very intriguing.  It is quite interesting that we can 
extract such dynamical information from the sky.     

We have also found that in the case without chiral symmetry breaking, the 
compositeness scale should be very low ($\lsim \mev$) in order to have the 
signal.  Therefore, we have discussed possible collider constraints and 
signatures.  Surprisingly, and fortunately or unfortunately, if we take the 
parameters to be such that there are signals in the CMB, we find no
additional constraints on the parameters from terrestrial experiments.

\acknowledgments
I thank Nima Arkani-Hamed for introducing me to and educating me on 
composite neutrinos.  I also thank Lawrence Hall, Ken Lane, Markus Luty, 
Danny Marfatia, Aaron Pierce, Veronica Sanz, and Martin Schmaltz for useful 
conversations.  Finally, I thank Zackaria Chacko for pointing out a crucial 
error and also reading the manuscript with many fruitful comments.  This work 
is supported by the U.S. Department of Energy under contract 
DE-FG03-91ER-40676.

\appendix

\section*{Appendix: Models With Massless Fermionic Composites}
Here we summarize some relevant results from \cite{Dimopoulos:1980hn}.  
We discuss two cases, the one without ``pions'' and the one with ``pions''.

\subsection*{(a)~A Model Without Chiral Symmetry Breaking}
Consider an $SU(6)$ gauge theory with fermions, $\psi_{aI}$ and  
$\chi^{ab}=-\chi^{ba}$, where $I$ is a flavor index ($I=1,2$), while 
$a,b$ are gauge indices ($a,b=1,\cdots,6$).  The gauge index corresponds to 
the fundamental of $SU(6)$ if it is downstairs, or the anti-fundamental if it 
is upstairs.  This theory is asymptotically-free and has the following flavor 
symmetries: 
\begin{itemize}
 \item{An $SU(2)$ symmetry acting on the two $\psi$s.}
 \item{A $U(1)$ symmetry, under which $\psi$ and $\chi$ have charge $2$ and 
       $-1$, respectively.}  
 \item{A $Z_2$ symmetry, under which $\psi \to -\psi$.}
 \item{A $Z_4$ symmetry, under which $\chi \to i\chi$.} 
\end{itemize}
The principle of complementarity applies to this theory, and one can analyze 
the theory both in the Higgs picture and in the confinement picture to  
verify that they give consistent results.  For our purpose, we would like to 
describe the results in the confinement language.  At low energies beneath 
the confinement scale, we have 
three massless ``baryons'', $B_{IJ}=B_{JI}=\psi_{\{ I} (\psi_{J\} } \chi)$, 
which is a triplet of the flavor $SU(2)$ and has charge 3 under the flavor 
$U(1)$.  The baryons do not transform under the $Z_2$, while they do transform 
as $B \to iB$ under the $Z_4$.  All the flavor symmetries remain unbroken 
at low energies, and indeed one can check that all of continuous and discrete 
anomalies match.  Since the theory has no flavor symmetry breaking, there are 
no ``pions'', and therefore these three baryons are the only massless 
particles from this theory.

\subsection*{(b)~A Model With Chiral Symmetry Breaking}
Consider an $SU(7)$ gauge theory with fermions, $\psi^I_{abc}$ and 
$\chi_S^{ab}$, where $I$ and $S$ are flavor indices ($I=1,2,3$ and $S=1,2$), 
while $a,b,c$ are gauge indices ($a,b,c=1,\cdots,7$).  The gauge index 
corresponds to the fundamental of $SU(7)$ if it is downstairs, or the 
anti-fundamental if it is upstairs.  All gauge indices are 
anti-symmetrized.  This theory is asymptotically-free and has the following 
flavor symmetries:
\begin{itemize}
 \item{An $SU(3)$ symmetry acting on the three $\psi$s.}
 \item{An $SU(2)$ symmetry acting on the two $\chi$s.}
 \item{A $U(1)$ symmetry, under which $\psi$ and $\chi$ has charge $-1$ and 
       $3$, respectively.}
 \item{A $Z_{30}$ symmetry, under which $\psi \to \e^{i\pi/15} \psi$.}
 \item{A $Z_{10}$ symmetry, under which $\chi \to \e^{i\pi/5} \chi$.} 
\end{itemize} 
The principle of complementarity applies to this theory, and one can analyze 
the theory both in the Higgs picture and in the confinement picture to  
verify that they give consistent results.  For our purpose, we would like to 
describe the results in the confinement language.  At low energies beneath 
the confinement scale, the flavor symmetries are broken down to the $U(1)$ and 
$Z'_{30} \equiv Z_{30} Z_{10}^{-1}$.  We have three massless ``baryons'', and 
the baryon has charge 7 under the $U(1)$ and transforms as 
$B \to \e^{-i7\pi/15} B$ under the $Z'_{30}$.  More specifically, 
$B_{IS} = \epsilon_{IJK} \phi^{*J}_a \phi^{*K}_b \chi^{ab}_S$, where 
$\phi^a_I= \epsilon_{IJK} \epsilon^{abcdefg} \psi^J_{bcd} \psi^K_{efg}$.  
Although it appears here that there are $3\times 2=6$ states of $B$ rather 
than 3, one can show that 3 of them actually marry with the 3 states of 
$B'^I = \epsilon^{JKL} \phi^a_J \phi^b_K \phi^c_L \psi^I_{abc}$ to become 
massive.  Indeed, such masses respect the low-energy flavor symmetries, 
because $B'$ has charge $-7$ under the $U(1)$ and transforms as 
$B' \to e^{i7\pi/15} B'$ under the $Z'_{30}$.   The leftover, three states 
of $B$ remain completely massless.  One can verify that all of the anomalies 
for the $U(1)$ and the $Z'_{30}$ match, 
while the anomalies involving other symmetries do not match.  This is in 
accord with the results from the Higgs picture analysis where there are just
three massless fermions with the same quantum number as $B$ and all the 
flavor symmetries except the $U(1)$ and the $Z'_{30}$ are broken 
by condensates.  Therefore, we conclude that the flavor symmetries are  
partially broken in this theory, and in particular the breaking 
of $SU(3)\times SU(2)\times U(1)$ down to $U(1)$ produces 11 massless 
``pions''.


\begin{thebibliography}{99}

\bibitem{Chacko:2003dt}
Z.~Chacko, L.~J.~Hall, T.~Okui and S.~J.~Oliver,
``CMB signals of neutrino mass generation,''
arXiv:hep-ph/0312267.

\bibitem{Arkani-Hamed:1998pf}
N.~Arkani-Hamed and Y.~Grossman,
``Light active and sterile neutrinos from compositeness,''
Phys.\ Lett.\ B {\bf 459}, 179 (1999)
[arXiv:hep-ph/9806223].

\bibitem{Fukuda:1998mi}
Y.~Fukuda {\it et al.}  [Super-Kamiokande Collaboration],
``Evidence for oscillation of atmospheric neutrinos,''
Phys.\ Rev.\ Lett.\  {\bf 81}, 1562 (1998)
[arXiv:hep-ex/9807003]; 

\bibitem{Ahmad:2002jz}
Q.~R.~Ahmad {\it et al.}  [SNO Collaboration],
``Direct evidence for neutrino flavor transformation from neutral-current
interactions in the Sudbury Neutrino Observatory,''
Phys.\ Rev.\ Lett.\  {\bf 89}, 011301 (2002)
[arXiv:nucl-ex/0204008]. \\

\bibitem{Eguchi:2002dm}
K.~Eguchi {\it et al.}  [KamLAND Collaboration],
``First results from KamLAND: Evidence for reactor anti-neutrino
disappearance,''
Phys.\ Rev.\ Lett.\  {\bf 90}, 021802 (2003)
[arXiv:hep-ex/0212021].\\

\bibitem{Smy:2003jf}
M.~B.~Smy {\it et al.}  [Super-Kamiokande Collaboration],
``Precise measurement of the solar neutrino day/night and seasonal 
variation in Super-Kamiokande-I,''
Phys.\ Rev.\ D {\bf 69}, 011104 (2004)
[arXiv:hep-ex/0309011].

\bibitem{Gherghetta:2003he}
T.~Gherghetta,
``Dirac neutrino masses with Planck scale lepton number violation,''
Phys.\ Rev.\ Lett.\  {\bf 92}, 161601 (2004)
[arXiv:hep-ph/0312392].

\bibitem{Barger:2003zg}
V.~Barger, J.~P.~Kneller, H.~S.~Lee, D.~Marfatia and G.~Steigman,
``Effective number of neutrinos and baryon asymmetry from BBN and WMAP,''
Phys.\ Lett.\ B {\bf 566}, 8 (2003)
[arXiv:hep-ph/0305075].

\bibitem{Hu:1995fq}
W.~Hu, D.~Scott, N.~Sugiyama and M.~J.~White,
``The Effect of physical assumptions on the calculation of microwave 
background anisotropies,''
Phys.\ Rev.\ D {\bf 52}, 5498 (1995)
[arXiv:astro-ph/9505043].

\bibitem{Hannestad:2003xv}
  S.~Hannestad,
  JCAP {\bf 0305}, 004 (2003)
  [arXiv:astro-ph/0303076].

\bibitem{Bashinsky:2003tk}
S.~Bashinsky and U.~Seljak,
``Neutrino Perturbations in CMB Anisotropy and Matter Clustering,''
arXiv:astro-ph/0310198.

\bibitem{Hannestad:2004qu}
  S.~Hannestad,
  JCAP {\bf 0502}, 011 (2005)
  [arXiv:astro-ph/0411475].

\bibitem{Trotta:2004ty}
  R.~Trotta and A.~Melchiorri,
  arXiv:astro-ph/0412066.

\bibitem{Weinberg:1978kz}
S.~Weinberg,
``Phenomenological Lagrangians,''
PhysicaA {\bf 96}, 327 (1979).

\bibitem{Manohar:1983md}
A.~Manohar and H.~Georgi,
``Chiral Quarks And The Nonrelativistic Quark Model,''
Nucl.\ Phys.\ B {\bf 234}, 189 (1984); \\

\bibitem{Georgi:kw}
H.~Georgi,
``Weak Interactions And Modern Particle Theory,'' 
(Benjamin/cummings, Menlo Park, USA, 1984); \\

\bibitem{Georgi:1986kr}
H.~Georgi and L.~Randall,
``Flavor Conserving CP Violation In Invisible Axion Models,''
Nucl.\ Phys.\ B {\bf 276}, 241 (1986).

\bibitem{Spergel:2003cb}
D.~N.~Spergel {\it et al.}
``First Year Wilkinson Microwave Anisotropy Probe (WMAP) Observations:
Determination of Cosmological Parameters,''
Astrophys.\ J.\ Suppl.\  {\bf 148}, 175 (2003)
[arXiv:astro-ph/0302209]; \\

\bibitem{Tegmark:2003ud}
M.~Tegmark {\it et al.}  [SDSS Collaboration],
``Cosmological parameters from SDSS and WMAP,''
arXiv:astro-ph/0310723; \\

\bibitem{Hannestad:2003ye}
S.~Hannestad and G.~Raffelt,
``Cosmological mass limits on neutrinos, axions, and other light particles,''
arXiv:hep-ph/0312154.

\bibitem{Weinheimer:tn}
C.~Weinheimer {\it et al.},
``High Precision Measurement Of The Tritium Beta Spectrum Near Its  Endpoint
And Upper Limit On The Neutrino Mass,''
Phys.\ Lett.\ B {\bf 460}, 219 (1999); \\

\bibitem{Lobashev:tp}
V.~M.~Lobashev {\it et al.},
``Direct Search For Mass Of Neutrino And Anomaly In The Tritium
Beta-Spectrum,''
Phys.\ Lett.\ B {\bf 460}, 227 (1999); \\

\bibitem{Bonn:tw}
J.~Bonn {\it et al.},
``The Mainz Neutrino Mass Experiment,''
Nucl.\ Phys.\ Proc.\ Suppl.\  {\bf 91}, 273 (2001).

\bibitem{Beacom:2004yd}
J.~F.~Beacom, N.~F.~Bell and S.~Dodelson,
``Neutrinoless Universe,''
arXiv:astro-ph/0404585.

\bibitem{Dohmen:mp}
C.~Dohmen {\it et al.}  [SINDRUM II Collaboration.],
``Test Of Lepton Flavor Conservation In Mu $\to$ E Conversion On Titanium,''
Phys.\ Lett.\ B {\bf 317}, 631 (1993); \\

\bibitem{Ahmad:1988ur}
S.~Ahmad {\it et al.},
``Search For Muon - Electron And Muon - Positron Conversion,''
Phys.\ Rev.\ D {\bf 38}, 2102 (1988).

\bibitem{Hagiwara:fs}
K.~Hagiwara {\it et al.}  [Particle Data Group Collaboration],
``Review Of Particle Physics,''
Phys.\ Rev.\ D {\bf 66}, 010001 (2002).

\bibitem{Dimopoulos:1980hn}
S.~Dimopoulos, S.~Raby and L.~Susskind,
``Light Composite Fermions,''
Nucl.\ Phys.\ B {\bf 173}, 208 (1980).

\end{thebibliography}
\end{document}